\documentstyle[prl,aps,epsfig,preprint]{revtex}
\begin{document}
\tighten

\title{Quark Orbital-Angular-Momentum \\
        Distribution in the Nucleon}
\author{Pervez Hoodbhoy\footnote{Fulbright scholar, on leave from
Department of Physics, Quaid-e-Azam University, Islamabad 45320,
Pakistan.}, Xiangdong Ji and Wei Lu} \bigskip

\address{
Department of Physics \\
University of Maryland \\
College Park, Maryland 20742 \\
{~}}

\date{UMD PP\#98-119 ~~~DOE/ER/40762-149~~~ April 1998}

\maketitle

\begin{abstract}
We introduce gauge-invariant quark and gluon 
angular momentum distributions after making 
a generalization of the angular momentum density 
operators. From the quark angular momentum 
distribution, we define the gauge-invariant
and leading-twist quark {\it orbital} angular
momentum distribution $L_q(x)$. The latter 
can be extracted from data on the polarized and 
unpolarized quark distributions and the off-forward 
distribution $E(x)$ in the forward limit. 
We comment upon the evolution equations obeyed 
by this as well as other orbital distributions 
considered in the literature.
\end{abstract}
\pacs{xxxxxx}

\narrowtext

It has long been suspected that the quark orbital 
angular momentum (OAM) plays an important role in 
the structure of the nucleon, even though empirical evidence
suggests that the constituent quarks are 
predominantly in the $s$-wave state in the
naive quark model. For instance, Sehgal \cite{seh} 
noticed, following the work of Ellis and Jaffe \cite{ej},
that the quark spin falls short in accounting for 
the spin of the proton. He considered seriously 
the possibility of a sizable quark OAM filling 
in this gap. Ratcliffe studied the angular momentum 
conservation in the parton splitting processes, pointing out that
the OAM can be generated by parton
helicity during the scale evolution \cite{rat}. 
Since the EMC data on polarized deep-inelastic
scattering \cite{emc}, a large number of 
theoretical papers have appeared on the spin structure 
of the proton, some of which have directly and indirectly 
considered the issue of the quark OAM \cite{song}.

In Ref. \cite{ji1}, the evolution equation for the 
quark and gluon OAM was derived 
in the light-cone gauge and coordinates, 
and in leading-order QCD perturbation theory. 
Based on this, it was concluded that
the fraction of the nucleon spin carried by quarks is
about $\hbar/4$ in the asymptotically large 
scale limit. Taking into account the polarized
DIS data \cite{emc,dis}, this indicates that 
approximately $\hbar/8$ of the nucleon spin 
resides in the quark OAM in the asymptotic limit. 
In Ref. \cite{ji2}, a proposal was made to gain 
access experimentally to the quark OAM
at a finite scale $\mu^2$ by studying
virtual Compton scattering in the deep-inelastic
limit. 

Recently, Hagler and Sch\"afer \cite{hs}, and Harindranath 
and Kundu \cite{hk} studied the quark OAM 
distribution as a function of Feynman $x$ in 
the nucleon. They considered a tower of operators, 
\begin{equation}
   \bar \psi \gamma^+ (x^1i\partial^2- 
     x^2i\partial^1) iD^+\cdots iD^+ \psi \ , 
\end{equation}
whose matrix elements are used to define the moments 
of the quark OAM distribution. They derived the one-loop
evolution equation which has a
homogeneous term the same as the evolution kernel for 
the unpolarized quark distribution\cite{ap}, 
generalizing a result 
found earlier for the first moment \cite{ji1,ji2}. 
The mixing of the quark OAM distribution 
with the quark helicity distribution 
can also be related to the polarized and unpolarized 
Altarelli-Parisi kernels, as elucidated recently 
by Teryaev \cite{ter}. 

In this paper, we introduce an alternative
quark OAM distribution. Our distribution is motivated from 
the quark angular momentum distribution
defined from the form factors or 
matrix elements of spin-independent 
twist-two operators. The new quark OAM distribution 
can be extracted from the polarized and unpolarized
quark distributions and the off-forward
distributions $E(x)$ introduced in Ref. \cite{ji2}. 
One important feature of the new distribution is 
that it is leading twist and its evolution is completely 
determined by that of polarized and unpolarized quark and gluon 
distributions to all orders in perturbation theory.  

We start with the familiar spin-independent, 
twist-two, flavor-singlet (summing over 
quark flavors) quark operators, 
\begin{equation}
   O_q^{\beta\mu_1\cdots\mu_n}(\xi) = 
     \bar \psi \gamma^{(\beta} 
   \stackrel{\leftrightarrow}{iD^{\mu_1}}...
       \stackrel{\leftrightarrow}{iD^{\mu_n)}} \psi(\xi) \ , 
\end{equation} 
where all indices $\beta,$ $\mu_1,$ ..., $\mu_n$ 
are symmetric and all traces have been 
subtracted, as indicated by the bracket $(\cdots)$. Note that the $n=1$
operator is just the traceless part of the 
energy-momentum tensor of quarks. From
$O_q^{\beta\mu_1\cdots\mu_n}$,  we define a new tower of 
twist-two operators which generalize the angular 
momentum density, 
\begin{equation}
    M_q^{\alpha\beta\mu_1\cdots\mu_n}(\xi)
    = \xi^\alpha O_q^{\beta\mu_1\cdots\mu_n}(\xi) 
      - \xi^\beta O_q^{\alpha\mu_1\cdots\mu_n}(\xi)
      - ({\rm traces})\ . 
\end{equation}
where the subtraction of traces ensures that 
$M_q^{\alpha\beta\mu_1\cdots\mu_n}$ belongs to 
an irreducible representation of the Lorentz group. 
The matrix element of the above operator 
in the nucleon state $|PS\rangle$, where
$P$ and $S$ are respectively the momentum 
and polarization vectors ($P^2=M^2$, $S^2=-M^2$, 
$S\cdot P =0$), is
\begin{eqnarray}
\left\langle PS \left|\int d^4\xi
M_q^{\alpha\beta\mu_1\cdots\mu_n}(\xi)\right|PS
\right\rangle &= &2J_{qn} {2S_\rho P_\sigma\over (n+1)M^2}
  \left(\epsilon^{\alpha\beta\rho\sigma} P^{\mu_1}\cdots P^{\mu_n} 
+ \epsilon^{[\alpha\mu_1\rho\sigma} P^{\beta]}\cdots P^{\mu_n} 
\right. \nonumber \\
 && \left.+ \cdots 
+ \epsilon^{[\alpha\mu_n\rho\sigma} P^{\mu_1}\cdots P^{\beta]} 
- ({\rm traces}) \right) (2\pi)^4\delta^4(0) \ , 
\label{me}
\end{eqnarray}
where $[\alpha\beta]$ indicates antisymmetrization of
the two indices. Here terms with derivatives on the
$\delta$ function have been omitted because they depend on the 
center-of-mass motion of the nucleon. The matrix element 
$J_{qn}$ generalizes the nucleon spin 
carried by quarks $J_{q1}$. In fact, we can define from it 
a quark angular momentum distribution $J_q(x)$ such that
\begin{equation}
   \int^1_{-1} J_q(x) x^{n-1} dx = J_{qn} \ . 
\end{equation}
$J_q(x)$ at negative $x$ represents the antiquark
contribution. If we limit the support to $0<x<1$, the angular
momentum distribution is $J_q(x) + J_q(-x)$. 
Obviously one can extend all of the above 
discussions to gluons and define the gluon 
angular momentum distribution $J_g(x)$. 

The matrix element $J_{qn}$, or equivalently,
the angular momentum distribution $J_q(x)$
can be obtained from the off-forward distribution
$H(x,\xi,t)$ and $E(x, \xi, t)$ in the forward
limit $\xi=t=0$. According to Ref. \cite{ji2}, 
these distributions are defined in terms of the off-forward
matrix element of the light-cone string operator,
\begin{eqnarray}
 &&  \int {d\lambda\over 2\pi} e^{i\lambda x}
    \left\langle P'\left|\overline \psi 
\left(-{\lambda \over 2}n\right)
      \not \! n {\cal P}e^{-ig\int^{-\lambda/2}_{\lambda/ 2}
       d\alpha ~n\cdot A(\alpha n)} 
    \psi\left({\lambda \over 2}n\right) \right| P\right\rangle 
    \nonumber \\
  = && H_q(x, \xi, t)~ \overline U(P')\not\! n U(P)
    + E_q(x, \xi, t)~ \overline U(P') {i\sigma^{\mu\nu}
  n_\mu \Delta_\nu \over 2M} U(P) \ . 
\label{string}
\end{eqnarray}  
where ${\overline U}(P')$ and $U(P)$ are the Dirac spinors,
$\Delta = P'-P$, $\overline P = (P'+P)/2$, and $t=\Delta^2$.
Vector $n$ along the light-cone ($n^2=0$) is 
conjugate to $\overline P$ in the sense that $\overline 
P\cdot n=1$. Taking the $n$-th moments 
of the above expression, we get,
\begin{eqnarray}
 && n_{\mu_1}n_{\mu_2}\cdots n_{\mu_n} \langle P'
     |O^{\mu_1\mu_2\cdots
   \mu_n}_q|P\rangle  \nonumber \\
 = && H_{qn}(\xi, t) \overline U(P')\not\! n U(P)
     + E_{qn}(\xi, t) \overline U(P') {i\sigma^{\mu\nu}
  n_\mu \Delta_\nu \over 2M} U(P) \ ,
\label{contr}  
\end{eqnarray}
where
\begin{eqnarray}
   H_{qn}(\xi, t) &=& \int^1_{-1} dx x^{n-1} H_q(x, \xi, t) dx \ , 
\nonumber \\
   E_{qn}(\xi, t) &=& \int^1_{-1} dx x^{n-1} E_q(x, \xi, t) dx \ , 
\end{eqnarray} 
are the moments of the off-forward distributions. 
To find their relation with $J_{qn}$, we write down 
all possible elastic form factors of the twist-two operators
after using the constraints from Lorentz symmetry 
and parity and time reversal invariance
\begin{eqnarray}
\langle P'| O^{\mu_1\cdots \mu_n}_q |P\rangle
   &= &{\overline U}(P') \gamma^{(\mu_1} U(P) 
 \sum_{i=0}^{[{n-1\over 2}]}
       A_{qn,2i}(t) \Delta^{\mu_2}\cdots \Delta^{\mu_{2i+1}}
      \overline{P}^{\mu_{2i+2}}\cdots
      \overline{P}^{\mu_n)}  \nonumber \\
    &&  + ~  {\overline U}(P'){\sigma^{(\mu_1\alpha}
     i\Delta_\alpha \over 2M}U(P)   \sum_{i=0}^{[{n-1\over 2}]}
     B_{qn,2i}(t) \Delta^{\mu_2}\cdots \Delta^{\mu_{2i+1}}
      \overline{P}^{\mu_{2i+2}}\cdots
    \overline{P}^{\mu_n)} \nonumber \\
     &&  + ~ C_{qn}(t) ~{\rm Mod}(n+1,2)~{1\over M}\bar U(P') U(P)
    \Delta^{(\mu_1} \cdots \Delta^{\mu_n)} \ ,
\label{form}
\end{eqnarray}              
For $n\ge 1$, even or odd, there are $n+1$
form factors. $C_n(t)$ is present only when $n$ is even.      
Substituting the above expression into Eq. (\ref{me}),
we have
\begin{equation}
    J_{qn} = {1\over 2}\Big(A_{qn+1,0}(0)  + B_{qn+1,0}(0)\Big)\ . 
\end{equation}
Contracting both sides of Eq. (\ref{form}) with
$n^{\mu_1}\cdots n^{\mu_n}$ and comparing the result
with Eq. (\ref{contr}), we obtain the desired relation
\begin{eqnarray}
    J_{qn} = {1\over 2}\Big(H_{qn+1}(0,0) + E_{qn+1}(0,0)\Big) \ . 
\end{eqnarray}
Clearly $H_q(x, 0, 0)$ is just the singlet quark
distribution $q(x)$. For convenience we abbreviate 
$E_q(x,0,0)$ as $E_q(x)$. Then the above relation 
can be translated to one for the quark angular 
momentum distribution,
\begin{equation}
    J_q(x) = {1\over 2}x\Big(q(x) + E_q(x)\Big) \ .
\label{rela}
\end{equation}
The distribution $q(x)$ has been extracted from 
high-energy scattering with good accuracy
\cite{cteq}. However, $E(x)$ is 
unknown at present except for its first moment 
(see below). 
 
Since forming spatial moments of an operator 
as in Eq. (\ref{me}) does not change its short 
distance behavior, the evolution equation 
for the angular momentum distributions
$J_q(x)$ and $J_g(x)$ is exactly the same as 
that for the unpolarized quark and gluon
distributions,
\begin{equation}
   {d\over d\ln \mu^2}
           \left( \begin{array}{c} J_{nq}(\mu) \\ J_{ng}(\mu )
     \end{array} \right)
  = \left(\begin{array}{cc} \gamma_{qq}(n+1) & \gamma_{qg}(n+1) \\
                            \gamma_{gq}(n+1) & \gamma_{gg}(n+1) 
                      \end{array} \right)
            \left( \begin{array}{c} J_{nq}(\mu) \\ J_{ng}(\mu)
      \end{array} \right)  
\label{evol}    
\end{equation}  
where the anomalous dimension matrix is a perturbation
series in $\alpha_s$ and the leading order result can be found 
in Ref. \cite{ap}. Of course, according to Eq. (\ref{rela})
the same conclusion follows from 
the evolution of the off-forward parton distributions
\cite{ji2}.

Now we can define the quark orbital angular momentum 
(OAM) distribution by looking at the 
structure of $M_q^{\alpha\beta\mu_1\cdots\mu_n}$.  Since it
is already an irreducible operator under 
Lorentz transformations, the generalized OAM 
operator that it contains is not irreducible and
hence the nucleon matrix element of the latter 
cannot be a Lorentz scalar. This is an intrinsic limitation
to the significance of the notion of the orbital angular
momentum in a relativistic quantum theory. 
However, if we are limited to a class of coordinates 
in which the nucleon has a definite helicity, 
we still can define the orbital 
angular momentum and its distribution in Feynman $x$ 
\cite{ji3}. In the following discussion,
we assume the nucleon is moving in the $z$ direction
and is polarized in the helicity +1/2 state.
We consider only the specific components
$\alpha=1,\beta=2,\mu_1=\cdots=\mu_n=+$ of the 
$M_q$ operator and take its spatial integral,
\begin{equation}
    \int d^3\xi M_q^{12+\cdots+} 
        = \int d^3\xi \left(\xi^1 M_q^{(2+\cdots+)}
          - \xi^2 M_q^{(1+\cdots+)}\right) \ , 
\end{equation} 
where we have introduced the light-cone coordinates:
$\xi^\pm = (\xi^0\pm \xi^3)/\sqrt{2}$. The spatial integral 
element is $d^3\xi = d^2\xi_\perp d\xi^-$. One can perform 
partial integrations to make all derivatives act
on the right,
\begin{eqnarray} 
    \int d^3\xi M_q^{12+\cdots+}
   &=& {1\over n+1} \int d^3\xi \xi^1 \left[
     \bar \psi \gamma^2 iD^+\cdots iD^+ \psi \right. \nonumber \\ 
   && \left.  +~ \bar \psi \gamma^+ iD^2\cdots iD^+ \psi
    ~+~ \cdots ~ + ~\bar \psi \gamma^+ iD^+ \cdots iD^2\psi  
    \right] - (1\leftrightarrow 2) \ . 
\label{101}
\end{eqnarray}
The first term in the bracket can be further manipulated, 
\begin{eqnarray}
     \bar \psi \gamma^2 iD^+\cdots iD^+ \psi
    &=& {1\over n}\left[ \bar \psi(\gamma^2 iD^+ 
     - \gamma^+ iD^2)iD^+\cdots iD^+\psi ~+ ~\cdots 
     \right.\nonumber \\
    && \left.  +~ \bar \psi iD^+\cdots iD^+
      (\gamma^2 iD^+ - \gamma^+ iD^2)\psi\right] \nonumber \\
    && ~+~ {1\over n}\left[ \bar \psi\gamma^+ 
      iD^2\cdots iD^+ \psi
    ~+~ \cdots ~+~\bar \psi \gamma^+ iD^+ \cdots iD^2\psi \right]
    \ . 
\end{eqnarray}
The second term in brackets on the right hand side is the same
as the remaining terms in Eq. (\ref{101}). So we concentrate
on the first term in the above equation. Using the following
identity, 
\begin{equation}
       \gamma^2 iD^+ - \gamma^+ iD^2
     = {1\over 2}\left[i\! \not\!\! D \gamma^+\gamma^2
        - \gamma^+\gamma^2 i\! \not\!\! D\right] \ , 
\end{equation}
and partially integrate the first derivative term, we have,
\begin{eqnarray} 
&& \int d^3\xi \left[ \bar \psi(\gamma^2 iD^+
     - \gamma^+ iD^2)iD^+\cdots iD^+\psi ~+~ \cdots
      ~+~ \bar \psi iD^+\cdots iD^+
      (\gamma^2 iD^+ - \gamma^+ iD^2)\psi\right] \nonumber \\
  = &&-i{n\over 2} \int d^3\xi\bar \psi \gamma^1\gamma^+\gamma^2
         iD^+\cdots iD^+ \psi 
     ~+~ \int d^3\xi \Big[\bar \psi\gamma^+(x^1\gamma^2)(igF^{\rho+})
             \gamma_\rho iD^+\cdots iD^+\psi \nonumber \\
  &&   ~+~\cdots ~+~ (n-1) \bar \psi\gamma^+iD^+\cdots iD^+
        (x^1\gamma^2)(igF^{\rho+}) \gamma_\rho \psi \Big] \ , 
\end{eqnarray}
where we have used the quark equations of motion to eliminate
some terms. 

Putting all the pieces together, we have
\begin{equation}
    \int d^3\xi M_q^{12+\cdots+}
        = S^{+\cdots +}_q
        + \tilde L^{+\cdots +}_q
        + \Delta L^{+\cdots+}_q \ , 
\end{equation}    
where,
\begin{eqnarray}
    S^{+\cdots +}_q &=& {2\over n+1} \int d^3\xi \bar 
       \psi \gamma^+ {\Sigma^3\over 2} iD^+ \cdots iD^+ \psi \ , 
\nonumber
\\
    \tilde L^{+\cdots+}_q &=& {1\over n} 
           \int d^3 \xi\left[\bar \psi \gamma^+ (x^1iD^2-x^2iD^1)
          iD^+\cdots iD^+) \psi \right. \nonumber \\
        &&  \left.~+ ~\cdots ~+ ~ \bar \psi \gamma^+ iD^+\cdots 
          iD^+ (x^1iD^2-x^2iD^1)\psi \right] \ , \nonumber \\
    \Delta L^{+\cdots + }_q
     & = & {1\over n(n+1)} \int d^3\xi 
       \left[\bar \psi \gamma^+ (x^1\gamma^2-x^2\gamma^1) (igF^{\rho+}
        \gamma_\rho)iD^+\cdots iD^+ \psi \right.\nonumber \\
     && \left. ~+~\cdots~+~ \bar \psi \gamma^+iD^+\cdots iD^+(x^1\gamma^2
       -x^2\gamma^1) (igF^{\rho+} \gamma_\rho) \psi\right] \ .   
\end{eqnarray}
The matrix element of $S^{+\cdots+}_q$ is clearly related to the polarized
quark distribution $\Delta q(x) $ which has been the major focus
of the polarized deep-inelastic scattering experiments in the 
last ten years \cite{emc,dis}. $\tilde L^{+\cdots+}_q$ seems to be a 
natural gauge-invariant generalization of the OAM operator. However,
from its definition, it is clear that 
$\tilde L^{+\cdots+}_q$ contains not only the leading twist but also
the higher-twist contributions. In particular, under a
change of renormalization scale, it mixes with $\Delta
L^{+\cdots+}_q$.  Only the sum $\tilde L^{+...+}_q+\Delta L^{+\cdots+}_q$ 
evolves as a leading-twist operator. Therefore, we define
the leading-twist, gauge-invariant generalization 
of the OAM operator as
\begin{equation}
    L^{+\cdots +}_q = \tilde L^{+\cdots +}_q + \Delta L^{+\cdots +} \ . 
\end{equation}
Its matrix element in the nucleon state is,
\begin{equation} 
   \langle PS|L^{+\cdots +}_q|PS\rangle
    = {2L_{qn} \over n+1} 2P^+\cdots P^+ (2\pi)^3 \delta^3(0) \ .
\end{equation}
Let us emphasize again that the simple structure of the matrix
element is only possible in the specific class of frames that we
defined earlier. From the above, the OAM 
distribution function $L_q(x)$ can be defined as, 
\begin{equation} 
   \int^1_{-1} L_q(x) x^{n-1} dx = L_{qn}  \ , 
\end{equation}
which can be expressed in terms of the 
quark angular momentum and helicity distributions,
\begin{eqnarray}
     L_q(x) = J_q(x) - {1\over 2}\Delta q(x) \ . 
\label{od}
\end{eqnarray}
The definition of the matrix elements of 
$S^{+\cdots+}$ and its relation to the quark helicity
distribution is standard,
\begin{eqnarray}
\langle PS|S^{+\cdots +}_q|PS\rangle
  &=& {\Sigma_{n}\over n+1}2P^+\cdots P^+ (2\pi)^3 \delta^3(0)
\nonumber \\
  \int^1_{-1} dx x^{n-1} \Delta q(x) &=& \Sigma_{n} \ . 
\end{eqnarray}
Combining Eqs. (\ref{rela},\ref{od}), we have, 
\begin{equation}
      L_q(x) = {1\over 2}\Big[ x\Big(q(x) 
     + E_q(x)\Big) -\Delta q(x)\Big] \ . 
\end{equation}
Thus, the quark OAM distribution
can be determined entirely from the polarized and unpolarized
singlet quark distribution and the off-forward
distribution $E(x)$. 

The evolution equation for $L_{qn}(\mu)$ is straightforward.  
The evolution of $\Sigma_{n}$ is well-known,
\begin{equation}
     {d\over d\ln \mu^2} \Sigma_{n}
        = \Delta\gamma_{qq}(n) \Sigma_{n} 
     + \Delta\gamma_{qg}(n) \Delta g_n  \ , 
\end{equation}
where $\Delta g_n$ are the moments of the gluon
helicity distribution and the anomalous dimensions
can be found in Ref. \cite{ap}. Combining the above 
with Eq. (\ref{evol}), we have the evolution
equation for $L_{qn}$,
\begin{eqnarray}
    {d\over d\ln \mu^2} L_{qn}
     &=&   \gamma_{qq}(n+1)L_{qn}
      + {1\over 2}(\gamma_{qq}(n+1) - \Delta\gamma_{qq}(n))
        \Sigma_{n} \nonumber \\
      && + \gamma_{qg}(n+1)J_{gn} - 
      {1\over 2}\Delta \gamma_{qg}(n)\Delta g_n \ , 
\label{le}
\end{eqnarray}
which is again valid to all orders in perturbation 
theory.

In view of the above discussion, we now have a new perspective
on the evolution equations obtained by Hagler and 
Sch\"afer, and Harindranath and Kundu. The quark 
OAM operators defined in these works (we will call it
$L^{+\cdots +}_{q0}$) differ from our
$L^{+\cdots +}_q$ by terms which depend on transverse 
polarizations of the gluon potentials ($A^1, A^2$).
Our calculation indicates that these extra terms do 
not mix in the $L^{+\cdots+}_{q0}$ operator 
at the one-loop level, although at higher-loops they must 
do so that $L^{+\cdots+}_q$ maintain an all-order simple 
evolution in Eq. (\ref{le}). Therefore at the one-loop level,
$L^{+\cdots +}_{q0}$ and $L^{+\cdots +}_q$ 
have an identical homogeneous part in their 
evolution equation. Taking into account the extra factor
of $2\over n+1$ in our definition of the OAM matrix
elements, we deduce that $L_{qn}^0 + \Delta \Sigma_n/(n+1)$
has the same homogeneous evolution as $J_{qn}$ 
at one-loop level. Similar analysis for the
gluon angular momentum shows that $\tilde L_{gn}^0 + 2
\Delta g_n/(n+1)$ has the same homogeneous evolution as
$J_{gn}$ at one loop. From these, we immediately deduce
that 
\begin{eqnarray}
{d \over d \ln \mu^2}
  && \left(\begin{array}{c}
         L_{qn}^0(\mu) \\
          L_{gn}^0(\mu)
    \end{array} \right)
   = {\alpha_s(\mu)\over 2\pi}
    \left( \begin{array}{rr}
        \gamma_{qq}^0(n+1)  & \gamma^0_{qg}(n+1)  \\
        \gamma_{gq}^0(n+1) & \gamma^0_{gg}(n+1)  \\
      \end{array} \right)
        \left( \begin{array}{c}  
           L_{qn}^0(\mu) \\
           L_{gn}^0(\mu)
        \end{array} \right)   \nonumber \\
 + && {\alpha_s(\mu)\over 2\pi}
    {1\over n+1} \left( \begin{array}{rr}
        \gamma_{qq}^0(n+1)-\Delta \gamma_{qq}^0(n)  & 
       2\gamma_{qg}^0(n+1)-\Delta \gamma_{qg}^0(n)  \\
        \gamma_{gq}^0(n+1)-2\Delta \gamma_{gq}^0(n) 
      & 2\gamma_{gg}^0(n+1)-2\Delta\gamma_{gg}^0(n)  \\
      \end{array} \right)
        \left( \begin{array}{c}  
           \Delta \Sigma_{n}(\mu) \\
           \Delta g_{n}(\mu)
        \end{array} \right) + \cdots 
\end{eqnarray}
at one loop. The explicit terms agree with 
the known result \cite{hs,hk}. The ellipses denote 
the matrix elements of operators with explicit transverse
gluon fields which are present even at the one-loop level.
Because of these operators, the evolution equation
is not guaranteed to have a similar
simple structure at higher loops.

To determine the quark OAM distribution, 
one has to measure $E(x)$ from off-forward
hard scattering processes. Deeply-virtual Compton 
scattering has been proposed as a way to measure it 
\cite{ji2}. High-energy
diffractive vector-meson production can also 
provide access
to the distribution \cite{rad}. 
The first moment of $E(x)$ is related to the anomalous
magnetic moment,
\begin{equation}
  \int^1_{-1} E(x) dx = \kappa_u + \kappa_d + \kappa_s
\end{equation}
The experimental data on this is $0.33 \pm 0.39$, 
a fairly small number \cite{sample}. 
This could indicate a small size of $E(x)$. 
If $E(x)$ is indeed small, $L_q(x)$
can be determined entirely in terms of
$\Delta q(x)$ and $q(x)$. However, there is no
sound theoretical reason to neglect $E(x)$ at 
present. 

\acknowledgements 
PH thanks the Fulbright Foundation for sponsoring his
visit to the University of Maryland. This work is supported in part by
funds provided by the U.S.  Department of Energy (D.O.E.) under
cooperative agreement DOE-FG02-93ER-40762.

\end{document}